\documentclass[twocolumn,secnumarabic,amssymb, nobibnotes, aps, prd]{revtex4}
\usepackage{graphicx}

\begin{document}
\title{It is needed to shout "wake up".}
\author{A.V.  Nikulov}
%\email[]{nikulov@ipmt-hpm.ac.ru}
\affiliation{Institute of Microelectronics Technology and High Purity Materials, Russian Academy of Sciences, 142432 Chernogolovka, Moscow District, RUSSIA.} %nikulov@ipmt-hpm.ac.ru
%\date{}
\begin{abstract} It is manifested that the disregard of the epistemological problems connected with the fundamental obscurity in quantum mechanics results to the misunderstanding of the essence of this positivistic theory and to both funny and grandiose mistakes of contemporary publications. Some examples of these mistakes are considered.
 \end{abstract}

\maketitle

\narrowtext

The immense progress of physics and engineering of the XX century developed thanks to quantum theory is beyond any doubt. But John Bell in his Introductory remarks at Naples-Amalfi meeting, May 7, 1984 "Speakable and unspeakable in quantum mechanics" \cite{Bell1984} stated that {\it "This progress is made in spite of the fundamental obscurity in quantum mechanics. Our theorists stride through that obscurity unimpeded... sleepwalking?"} He said further \cite{Bell1984}: {\it "The progress so made is immensely impressive. If it is made by sleepwalkers, is it wise to shout 'wake up'? I am not sure that it is. So I speak now in a very low voice"}. But now it is needed to shout 'wake up'. Numerous contemporary publications in Physical Review Letters and other respectable journals demonstrate that the "sleepwalkers" may be liable to make both funny and grandiose mistakes. In the beginning of this articles an example of a funny mistake will be considered. It will be shown that this mistake is a consequence of misunderstanding by the authors of the subject of quantum-mechanical description. The difference of the subject of the orthodox quantum mechanics from the one of other theories of physics was debated heatedly by the founders of quantum theory. But {\it the wonderful difficulties of quantum mechanics were largely trivialized, swept aside as unimportant philosophical distractions} \cite{QuCh2005} by the most physicists of the posterior generations. Therefore many modern authors can not understand even the reason of the philosophical debate between the founders of quantum mechanics about its fundamental obscurity. 

The philosophical debate between the founders has resulted in particular to the revelation of non-locality of quantum mechanics. One of the manifestations of this non-locality proposed by Elitzur and Vaidman \cite{Elitzur} has provoked blind imitation \cite{Paraoanu,Strambini}. The mistake of the work \cite{Paraoanu} is a consequence of a mass delusion concerning the problem of superposition of macroscopically distinct quantum states. The connection of this delusion with the fundamental obscurity in quantum mechanics will be considered below. The badness of the theoretical result \cite{Paraoanu} must be clear for each physicist because of the contradiction with the fundamental law of momentum conservation. Using blindly the orthodox quantum formalism the authors \cite{Paraoanu} come to the conclusion that the probability of electron transmission through two arms of an Aharonov-Bohm ring should depend on magnetic flux value $\Phi $ inside this ring. According to the result shown on Fig.2 \cite{Paraoanu} all electrons should reflected at the $\Phi $ value equal half of the flux quantum $\Phi_{0} = 2\pi \hbar/e$. This false prediction of the non-local force-free momentum transfer from $p$ to $-p$ contradicting impermissibly to the conservation law is a consequence of the illusion that the orthodox quantum mechanics can describe a process of electron transmission. 

Heisenberg warned against this misunderstanding of quantum mechanics as a description of a real process: {\it "A real difficulty in the understanding of the Copenhagen interpretation arises, however, when one asks the famous question: But what happens 'really' in an atomic event?"} \cite{Heisenberg1958}. Impossibility of realistic description of some quantum phenomena Heisenberg demonstrated in particular \cite{Heisenberg1958} on the example of the double-slit interference experiment. We can easy describe the interference pattern 
$$P(y) = A_{1}^{2} + A_{2}^{2} + 2A_{1}A_{2} \cos(\Delta \varphi _{1} - \Delta \varphi _{2}) \eqno{(1)} $$ 
observed on a detecting screen at the same velocity $v$ of particles using superposition $\Psi = \Psi _{1} + \Psi _{2}$ of two probability functions $\Psi _{1} = A_{1}e^{i\varphi _{1}}$, $\Psi _{2}= A_{2}e^{i\varphi _{2}}$ of momentum $p = mv$ eigenstates describing two possible path $l_{1}$, $l_{2}$ through the first slit $\Delta \varphi _{1} = \int_{S}^{y}dr_{1}(p/\hbar ) + tE/\hbar$ and the second slit $\Delta \varphi _{2} = \int_{S}^{y}dr_{2}(p/\hbar ) + t E/\hbar $ between a particle source $S$ and a point $y$ on the detecting screen. $A_{1}$, $A_{2}$ are the amplitudes of an arrival probability at the point $y$ of a particle passing through the first, second slit. But as Heisenberg noted: {\it This example shows clearly that the concept of the probability function does not allow a description of what happens between two observations. Any attempt to find such a description would lead to contradictions; this must mean that the term 'happens' is restricted to the observation} \cite{Heisenberg1958}. Indeed, the probabilities $|\Psi _{1}|^{2} = A_{1}^{2}$, $|\Psi _{1}|^{2} = A_{1}^{2}$ do not change in space $r$ and time $t$. Actually only a particle arrival on a detecting screen is observed in the real double-slit interference experiment \cite{QuCh2005}. Thanks to these observations we can know that $N$ electrons \cite{electron}, for example, have transmitted through two slits. But we can not know, according to the complementarity principle, through what slit exactly each electron transmitted in order the interference pattern can be observed \cite{QuCh2005}.

Aharonov and Bohm noted more than fifty years ago \cite{AB1959} that the phase difference $\Delta \varphi _{1} - \Delta \varphi _{2} = \int_{S}^{y}dr_{1}(p/\hbar ) - \int_{S}^{y}dr_{2}(p/\hbar ) = \int_{S}^{y}dr_{1}(mv/\hbar ) - \int_{S}^{y}dr_{1}(mv/\hbar ) + \oint dr (eA/\hbar ) =2\pi  (l_{1} - l_{2})/\lambda _{deB} + 2\pi \Phi /\Phi _{0}$ and consequently, according to (1), the interference pattern should shift with magnetic flux $\Phi $ because of the relation $p = mv + eA$ between canonical momentum $p$ and electron velocity $v$ in the presence of a magnetic vector potential $A$. $\lambda _{deB} = 2\pi \hbar /mv$ is the de Broglie wavelength. The Aharonov-Bohm effects is paradoxical because of the change with $\Phi $ of the probability of electron arrival at a point $y$ though the electron never enters the region containing the magnetic field (non-locality) and without any forces having acted on the electron \cite{QuCh2005}. But in contrast to the result \cite{Strambini} the Aharonov-Bohm effects does not concede, in accordance with the conservation law, a dependence of the transmission probability $P_{tr} = \int dy P(y) = \int dy (A_{1}^{2} + A_{2}^{2}) = 1$ on magnetic flux. 

The mistakes made in \cite{Paraoanu,Strambini} and other publications may be considered as a consequence of a great disparity between the essence of the orthodox quantum mechanics and its history of development and studying. The fundamental obscurity about which Bell said and the heat debates between the fathers of quantum theory ought be attributed rather to epistemology than to quantum theory. Einstein, Heisenberg, Bohr and others agreed that quantum mechanics is very successful in describing quantum phenomena, but no a quantum reality. Their disagreement about the completeness of quantum description was consequence of the difference of opinion on the goal of science. Bohr and Heisenberg advocated this completeness on the assumption of impossibility to describe an objective reality: {\it "There is no quantum world. There is only an abstract quantum physical description. It is wrong to think that the task of physics is to find out how Nature is"}  \cite{Bohr1958}. {\it "In classical physics science started from the belief - or should one say from the illusion? - that we could describe the world or at least parts of the world without any reference to ourselves"} \cite{Heisenberg1958}. Arguing against this positivism point of view of Heisenberg, Bohr and other adherents of the Copenhagen interpretation Einstein persisted that {\it "it must seem a mistake to permit theoretical description to be directly dependent upon acts of empirical assertions, as it seems to me to be intended [for example] in Bohr's principle of complementarity"} \cite{Einstein1949}.

The absence of an universal quantum world, stated by Bohr, presumes the absence of an universal description necessity. Nevertheless orthodox quantum mechanics, in spite of its positivism, developed and was misinterpreted among most physicists as an universal description of a quantum world. This misunderstanding of the quantum mechanics essence results to universally interpretation of wave functions used for description of different phenomena observed, for example, in the double-slit interference experiment, mesoscopic ring \cite{Strambini} and even superconductor \cite{Paraoanu}. Just this false propensity for universality results to funny mistake \cite{Strambini} and mass delusion concerning reality of superconducting quantum bits, see References in \cite{Paraoanu}. 

The concept of measurement must have fundamental importance in a description of phenomena, i.e. results of measurements (observations). But as Bell noted justly: {\it "The concept of 'measurement' becomes so fuzzy on reflection that it is quite surprising to have it appearing in physical theory at the most fundamental level"} \cite{Bell1987}. In fact, this concept is reduced to the words on a collapse \cite{Neumann1932} of the function $\Psi $ or a 'quantum jump' {\it "from the 'possible' to the 'actual' taking place during the act of observation"} \cite{Heisenberg1958}. This concept presumes a difference of quantum phenomena depending on presence or absence of the act of observation. The Aharonov-Bohm effects observed in the interference experiment \cite{AB1959} and mesoscopic ring demonstrate convincingly this difference \cite{FPP2008}.

Because of the requirement that any function must be single-valued $\Psi = |\Psi | e^{i\varphi } = |\Psi | e^{i(\varphi + 2\pi n )}$ at any point the phase difference in (1) should be divisible by $2\pi $ 
$$\Delta \varphi _{1} - \Delta \varphi _{2} = \oint_{l} dl \nabla \varphi = 2\pi n \eqno{(2)}$$ 
without the collapse at observation, for example, of electron arrival at a point $y$. The phase difference $\Delta \varphi _{1} - \Delta \varphi _{2} = \int_{S}^{y}dr_{1} \nabla \varphi - \int_{S}^{y}dr_{2} \nabla \varphi = \oint_{l} dr \nabla \varphi $ and the probability (1) can change uninterruptedly with the coordinate $y$ on the detecting screen and magnetic flux $\Phi $ only thanks to the collapse of the $\Psi$ function at the act of observation. The authors \cite{Strambini} did not take into account the absence of this act in the case of the Aharonov-Bohm ring. The requirement (2) valid in this case results to such well-known Aharonov-Bohm effect as the persistent current observed in normal metal \cite{Science09} and superconductor \cite{Science07} rings. The current of Cooper pairs can flow freely through the ring arms at any magnetic flux \cite{JETP07D} in defiance of the "prediction" \cite{Strambini} of zero transmission probability at $\Phi = 0.5\Phi _{0}$. The description of the periodicity in magnetic field \cite{JETP07D} connected with the Aharonov-Bohm effect is based on the requirement (2). Although the measurements reveal a paradoxical lack of correspondence between observed and predicted oscillations in magnetic field of the critical current of asymmetric rings \cite{JETP07D,JETP07J}.

Because of the history of the quantum mechanics development the same wave function seems to can be used for descriptions of both the double-slit interference experiment and superconductivity. This illusion results to numerous proposals to make quantum computer on base of superconductor structures, see References in \cite{Paraoanu}. There is important to remember that the basic principle of the idea of quantum computation, entanglement \cite{Schrod35} or Einstein-Podolsky-Rosen correlation \cite{EPR1935}, was introduced in 1935 by opponents of the Copenhagen interpretation in order to prove that quantum-mechanical description of physical reality can not be considered complete. Schrodinger introduced \cite{Schrod35} this principle as {\it "entanglement of our knowledge"} \cite{Entangl} because of its obvious contradiction with local realism. This contradiction results from the cardinal positive principle of the orthodox quantum mechanics, superposition of states, \cite{LandauL} and its instantaneous collapse at observation \cite{QuCh2005,EPR1935}. 

It is obvious that the Ginzburg-Landau wave function $\Psi _{GL} = |\Psi _{GL}| e^{i\varphi }$ describing the quite real density of the Cooper pairs $|\Psi_ {GL}|^{2} = n_{s}$ can not change because of the act of observation. There is important to remember that Schrodinger introduced wave function for a realistic description of quantum phenomena. Feynman in the Section "The Schrodinger Equation in a Classical Context: A Seminar on Superconductivity" of his Lectures on Physics \cite{FeynmanL} stated that Schrodinger {\it "imagined incorrectly that $|\Psi |^{2}$  was the electric charge density of the electron"}. Indeed, the positivistic interpretation by Born has allowed to evade inadmissible consequences at description of some phenomena observed on atomic level. But it was mistake to use this interpretation for all quantum phenomena including macroscopic one. This universal interpretation results to the illusion that each wave function can collapse and thus contradicts realism. Schrodinger \cite{Schrod35} as well as Einstein \cite{Einstein1949} used the term rather $\psi $ - function than wave function for the Schrodinger function in its positivistic interpretation. It is rational to use this term for the Schrodinger function which can collapse in order to distinguish it from the wave function, which, in accordance with its initial interpretation by Schrodinger \cite{Schrodinge26}, describes a real density and therefore can not collapse.

Using these different terms we should say that superconductivity is described with a wave function whereas quantum bit (qubit) must be described with a $\psi $ - function. Therefore authors of publications about superconducting qubits fabricate a $\psi $ - function in addition to the wave function describing superconducting state. The confidence of numerous authors that some phenomena observed at measurements of  superconducting qubits \cite{Clark08} give experimental evidence of superposition and its collapse is consequence of rather the development history than the essence of quantum mechanics. Some generations of physicists learned that the superposition of states is the cardinal positive principle of quantum mechanics \cite{LandauL}. Therefore most modern authors are inclined to assume superposition of states almost every quantum system, including macroscopic one, in spite of the contradiction even with macroscopic realism \cite{Leggett1985}. Mermin wrote as far back as 1985 \cite{Mermin1985} that {\it "In the question of whether there is some fundamental problem with quantum mechanics signaled by tests of Bell's inequality, physicists can be divided into a majority who are "indifferent" and a minority who are "bothered""}. The title "Is the moon there when nobody looks? Reality and the quantum theory" of the paper \cite{Mermin1985} indicates that the minority bother about the fundamental problem which bothered Einstein. A. Pais remembered in the book \cite{Pais} that Einstein asked he in 1950: {\it "Could you think that moon exists only when you look?"}

It is doubtful that many physicists can think seriously that the moon does not exist when nobody looks. But authors of numerous publications about flux qubit \cite{Clark08} repudiate the objective reality of macroscopic magnetic flux \cite{Leggett1985} with striking thoughtlessness, without solid grounds \cite{OI2009}, like the violation of the Bell's inequalities. The famous no-hidden-variables theorem (or, vulgarly, no-go theorem) \cite{Mermin1993} by John Bell \cite{Bell1964} proves that any realistic theory providing all predictions for the experiment outcomes given by the superposition principle should presume an inadmissible non-local interaction. Bell accentuated \cite{Bell1964} and demonstrated on the example of spin-1/2 \cite{Bell1966} that no-go theorem can not be correct without a separability or locality requirement, at least for a two-state quantum system. Therefore only from force of bad habit for the superposition principle numerous authors \cite{Paraoanu,Clark08} could believe that the phenomena without even a shadow of non-locality and even without the paradoxicality \cite{OI2009} of the Stern-Gerlach experiment testify against realism. The force of this habit is so strong that even the obvious contradiction of the assumption on flux qubit with the fundamental law of angular momentum conservation \cite{OI2009,Nat2009} is disregarded. 

The fantasy by the author \cite{Paraoanu} follows directly from the mass delusion that some macroscopic quantum phenomena observed at measurements of superconductor structures can not be described without the $\psi $ - function. The measurement considered in \cite{Elitzur} can be interaction-free only if it can not be described without the $\psi $ - function. The object can be detected in the way proposed in \cite{Elitzur} if light is only a wave. But the measurement could not be interaction-free in this case. The interaction-free measurement, non-locality, EPR correlation and other miracles could be conceivable only because of the wave-particle duality described by the $\psi $ - function. The duality is undoubtedly observed, for example, in the double-slit interference experiment \cite{QuCh2005,electron}. But it is mistake to think that the duality exists. Einstein, who introduced in 1905 light quanta (photons) and thus the duality, was forced to say on {\it "ghost waves (Gespensterfelder) guiding the photons"} \cite{Bohr1949} and confessed in 1951 that {\it "All these fifty years of conscious brooding have brought me no nearer to the answer to the question, 'What are light quanta?' Nowadays every Tom, Dick and Harry thinks he knows it, but he is mistaken"}, cited in \cite{Afshar07}. 

Einstein forewarned as far back as 1928 the sleepwalking about which Bell said in 1984 \cite{Bell1984}. He wrote to Schrodinger \cite{Lett1928}: {\it "The soothing philosophy-or religion?-of Heisenberg-Bohr is so cleverly concocted that it offers the believers a soft resting pillow from which they are not easily chased away"}, see the cite on the page 99 of \cite{QuCh2005}. The fundamental obscurity in quantum mechanics, according to Bell, may be connected with the Problem: {\it how exactly is the world to be divided into speakable apparatus...that we can talk about...and unspeakable quantum system that we can not talk about?} \cite{Bell1984}. Heisenberg forewarned also many times \cite{Heisenberg1958} that {\it "there is no way of describing what happens between two consecutive observations"} and {\it "that the concept of the probability function does not allow a description of what happens between two observations"}. Nevertheless numerous authors have no doubt about the ability of the orthodox quantum mechanics to describe the process of quantum computation which should be just between observations and use the $\psi$-function for the description of what happens between observations. The reason of this sleepwalking may be explain with the following remark by Einstein: {\it "Science without epistemology is - insofar as it is thinkable at all - primitive and muddled"} \cite{Einstein1949}. Just because of the neglect of the epistemological problems many physicist misinterpret  quantum mechanics as an universal theory of an objective reality. This misinterpretation results to both funny and grandiose mistakes of numerous publications.
   
\section*{Acknowledgement}
This work has been supported by a grant "Possible applications of new mesoscopic quantum effects for making of element basis of quantum computer, nanoelectronics and micro-system technic" of the Fundamental Research Program of ITCS department of RAS.

\end{document}